\documentclass[twocolumn,prl,preprintnumbers,nofootinbib,epsfig]{revtex4}
\usepackage{epsfig}% 
\usepackage{graphicx}% Include figure files
\usepackage{dcolumn}% Align table columns on decimal point
\usepackage{bm}% bold math

\newcommand{\bea}{\begin{eqnarray}}
\newcommand{\eea}{\end{eqnarray}}
\newcommand{\beq}{\begin{equation}}
\newcommand{\eeq}{\end{equation}}

\newcommand{\lsim}{~{}_{\textstyle\sim}^{\textstyle <}~}
\newcommand{\ba}{\begin{array}{c}}
\newcommand{\bat}{\begin{array}{cc}}
\newcommand{\ea}{\end{array}}

\def\slashchar#1{\setbox0=\hbox{$#1$}\dimen0=\wd0%
\setbox1=\hbox{/}\dimen1=\wd1%
\ifdim\dimen0>\dimen1%                     
\rlap{\hbox to
\dimen0{\hfil/\hfil}}#1\else                                     
\rlap{\hbox to \dimen1{\hfil$#1$\hfil}}/\fi}
\newcommand{\grts}{\stackrel{>}{_\sim}}

\newcommand{\beqa}{\begin{eqnarray}}
\newcommand{\eeqa}{\end{eqnarray}}
\begin{document}

%\twocolumn[\hsize\textwidth\columnwidth\hsize\csname
%@twocolumnfalse\endcsname

%\preprint{MAP-??}

\title{Neutrinoless Double Beta Decay and  
Lepton Flavor Violation 
}
%\thanks{  }
\author{V. Cirigliano $^1$} %\email{vincenzo@its.caltech.edu}
\author{A. Kurylov $^1$ } %
\author{M.J. Ramsey-Musolf $^{1,2,3}$}%
\author{P. Vogel $^1$}%
\affiliation{$^1$ 
Kellogg Radiation Laboratory, California Institute of Technology, 
Pasadena, CA 91125, USA 
} 
\affiliation{$^2$ Institute for Nuclear Theory, University of Washington, 
Seattle, WA 98195, USA}
\affiliation{$^3$ Department of Physics, University of Connecticut, Storrs, 
CT 06269, USA}

% \date{\today}% It is always \today, today,
             %  but any date may be explicitly specified

\begin{abstract}

We point out that extensions of the Standard Model with low scale
($\sim$ TeV) lepton number violation (LNV) generally lead to a pattern
of lepton flavor violation (LFV) experimentally distinguishable 
from the one implied by models
with GUT scale LNV.  As a consequence, muon LFV processes provide a
powerful diagnostic tool to determine whether or not the effective
neutrino mass can be deduced from the rate of neutrinoless double beta
decay. We discuss the role of $\mu \to e \gamma$ and
$\mu \to e$ conversion in nuclei, which will be studied with high
sensitivity in forthcoming experiments.

\end{abstract}

\pacs{Valid PACS appear here}% PACS, the Physics and Astronomy
                             % Classification Scheme.
%\keywords{Suggested keywords}%Use showkeys class option if keyword
                              %display desired
\maketitle

%\vspace{0.5cm}]
%\narrowtext

In the past few years convincing experimental evidence for neutrino
oscillations has been collected~\cite{nu-osc-expt}, implying that
neutrinos are massive particles, with masses much smaller than those
of other known fermions.
% with  $m_{\nu_i} \leq 2.2 \ {\rm eV} $~\cite{tritium}.  
Since the Standard Model of particle physics assumes that neutrinos
are massless, the study of neutrino mass and the properties of massive
neutrinos provides important clues about a more fundamental theory
that goes beyond the Standard Model.  Among the most urgent open
questions in neutrino physics are  the determination of (i) neutrino 
charge conjugation properties (Dirac or Majorana) and (ii) 
the absolute mass scale in the spectrum. 

The study of neutrinoless double beta decay ($0\nu\beta\beta$) can
help addressing these issues.  For one, observation of this $\Delta
L=2$ process would establish the existence of total lepton number
violation (LNV), thereby implying that neutrinos are massive Majorana
particles~\cite{sv82}. Ideally, the observation of $0\nu\beta\beta$
would also help determine the absolute scale of neutrino mass, since
the total decay rate ($\Gamma_{0 \nu \beta\beta}$) arising from light
Majorana neutrinos is proportional to the square of the effective
mass, $m_{\beta\beta}$ (defined precisely below). It has long been
recognized, however, that the extraction of $m_{\beta\beta}$ from
$\Gamma_{0 \nu \beta\beta}$ is problematic, since LNV interactions
involving heavy ($\sim$ TeV) particles can make comparably important
contributions to the rate. Thus, in the absence of additional
information about the mechanism responsible for $0\nu\beta\beta$, one
could not unambiguously infer $m_{\beta\beta}$ from $\Gamma_{0 \nu
\beta\beta}$.

In this Letter, we show that experimental searches for lepton {\em
flavor} violation (LFV) involving charged leptons can help to address
this problem by providing a powerful \lq\lq diagnostic tool" for
establishing the $0\nu\beta\beta$ mechanism. Here, we
focus on the Standard Model-forbidden processes
$\mu\to e\gamma$ and $\mu\to e$ conversion in nuclei that will be
studied with unprecedented sensitivity in the forthcoming
MEG~\cite{MEG} and MECO~\cite{MECO} experiments, respectively. 
The relevant branching ratios are $B_{\mu \rightarrow e \gamma}
= \Gamma (\mu \rightarrow e \gamma)/ \Gamma_\mu^{(0)}$ and $B_{\mu \to
e} = \Gamma_{\rm conv}/\Gamma_{\rm capt} $, where $\mu \to e \gamma$
is normalized to the standard muon decay rate $\Gamma_\mu^{(0)} =
(G_F^2 m_\mu^5)/(192 \pi^3)$, while $\mu \to e$ conversion is
normalized to the capture rate $\Gamma_{\rm capt}$. The new
experiments will probe $B_{\mu \rightarrow e \gamma} $ and $B_{\mu \to
e}$ at levels that would be sensitive to the effects of LFV induced by
interactions involving TeV scale particles.  
Since models for the generation of Majorana neutrino masses typically
also imply the existence of such interactions, studies of charged
lepton LFV can also provide insight into the mechanism of
$0\nu\beta\beta$. 

The main discriminating quantity in our analysis is the ratio ${\cal
R} = B_{\mu \to e}/B_{\mu \rightarrow e \gamma}$. We find that:

\begin{enumerate} 

\item 
Observation of both the LFV muon processes
$\mu \to e$ and $\mu \to e \gamma$ with relative ratio ${\cal R} \sim
10^{-2}$ implies, under generic conditions, that $\Gamma_{0 \nu \beta
\beta} \sim m_{\beta \beta}^2$.

\item 
On the other hand, observation of LFV muon processes with
relative ratio ${\cal R} \gg 10^{-2}$ could signal non-trivial LNV
dynamics at the TeV scale, whose effect on $0 \nu \beta \beta$ has to
be analyzed on a case by case basis. Therefore, in this scenario no
definite conclusion can be drawn based on LFV rates.
 
\item
Non-observation of LFV in muon processes in forthcoming 
experiments would imply either that the scale of non-trivial LFV and
LNV is  above a few TeV, and thus $\Gamma_{0 \nu \beta
\beta} \sim m_{\beta \beta}^2$, or that any TeV-scale LNV is
approximately flavor diagonal.

\end{enumerate}
(If only one process is observed, the deduced constraint on
${\cal R}$ may still be of use for our analysis.) 
Below, we explain the basis for the above observations and
discuss the requirements on scenarios that might circumvent them.

In general, $0 \nu \beta \beta$ can be generated by (i) light Majorana
$\nu$ exchange (helicity flip and non-flip) or (ii) heavy particle
exchange~ (see, e.g., \cite{heavy,Prezeau:2003xn}), resulting from LNV
dynamics at some scale above the electroweak one.  The relative size
of heavy ($A_H$) versus light particle ($A_L$) exchange contributions
to the decay amplitude can be crudely estimated as
follows~\cite{Mohapatra:1998ye}: 
\beq A_L \sim G_F^2 \frac{m_{\beta
\beta}}{ \bar{k}^2 } ,~ A_H \sim G_F^2 \frac{M_W^4}{\Lambda^5}
,~ \frac{A_H}{A_L} \sim \frac{M_W^4 \bar{k}^2  } {\Lambda^5
m_{\beta \beta} } \ ,
\label{eq:estimate}
\eeq 
where $m_{\beta \beta} = \sum_i U_{e i}^2 m_{\nu_i}$ is the effective
Majorana mass ($U_{\ell n}$ being the light neutrino mixing matrix),
$ \bar{k}^2 \sim ( 50 \ {\rm MeV} )^2 $ is the typical light
neutrino virtuality, and $\Lambda$ is the heavy scale relevant to the
LNV dynamics.  Therefore, $A_H/A_L \sim O(1)$ for $m_{\beta \beta}
\sim 0.1-0.5$ eV and $\Lambda \sim 1$ TeV, and thus the LNV dynamics
at the TeV scale leads to similar $0 \nu \beta \beta$ decay rate as
the exchange of light Majorana neutrinos with effective mass $m_{\beta
\beta} \sim 0.1-0.5$ eV.

By itself, the $0\nu \beta \beta$ lifetime measurement does not
provide the means for determining the underlying mechanism.  The
spin-flip and non-flip exchange can be, in principle, distinguished by
the measurement of the single-electron spectra or polarization (see
e.g. \cite{Doi}). In most cases, however, the observation of the
emitted electrons does not distinguish between light Majorana or heavy
particle exchange. Thus, one must look for phenomenological
consequences of the different mechanisms other than observables 
directly associated with $0\nu\beta\beta$.  Here we point out the link  
between $0 \nu \beta \beta$ mechanisms and muon LFV processes. 

It is useful to formulate the problem in terms of effective low energy
interactions obtained after integrating out the heavy degrees of
freedom that induce LNV and LFV dynamics.
Relevant quantities in this context are the LNV and LFV scales, which
in general may be distinct.  As long as both scales are well above the
weak scale, then $\Gamma_{0\nu \beta \beta} \sim m_{\beta \beta}^2$,
and one does not expect to observe LFV signals in forthcoming
experiments (item 3 above).  In scenarios with high scale LNV (but possibly
low scale LFV), such as SUSY-GUT~\cite{Barbieri:1995tw} or SUSY
see-saw~\cite{Borzumati:1986qx}, one expects again $\Gamma_{0\nu \beta
\beta} \sim m_{\beta \beta}^2$, as well as ${\cal R} \sim 10^{-2}$ 
(item 1).
The case where the scales of LNV and LFV are both relatively low
($\lsim$ TeV) is more subtle, as this scenario might lead to
observable LFV and at the same time to  ambiguities in interpreting a
positive signal in $0 \nu \beta \beta$. Here one needs to develop some
discriminating criteria, as we discuss in detail below. 

Denoting the new physics scale by $\Lambda$, the effective lagrangian
for $0 \nu \beta \beta$ has the form
\beq
{\cal L}_{0 \nu \beta \beta} = \displaystyle\sum_i \ 
\frac{\tilde{c}_i}{\Lambda^5}  \  \tilde{O}_i  
\qquad  \tilde{O}_{i} =  \bar{q} \Gamma_1 q \,  \
\bar{q} \Gamma_2 q \,   \bar{e} \Gamma_3 e^c   \ ,  
\label{eq:lag1}
\eeq
where we have suppressed the flavor and Dirac structures for
simplicity (a complete list of the dimension nine operators
$\tilde{O}_i$ can be found in Ref.~\cite{Prezeau:2003xn}).  For the
LFV interactions, one has
\beq
{\cal L}_{\rm LFV} = \displaystyle\sum_i \ 
\frac{c_i}{\Lambda^2}  \  O_i  \  , 
\label{eq:lag2}
\eeq
and a complete operator basis can be found in
Refs.~\cite{Raidal:1997hq,Kitano:2002mt}.  The LFV operators relevant to
our analysis are of the following type 
\bea 
O_{\sigma L} & = &   \displaystyle\frac{e}{(4 \pi)^2}
 \overline{\ell_{iL}} \, \sigma_{\mu \nu} 
i  
%\gamma 
\slashchar{D} \, \ell_{jL}  \  F^{\mu \nu}  + {\rm h.c.}  \ , 
\nonumber \\
O_{\ell L} & = &   \overline{\ell_{iL}} \, \ell^c_{jL} \  
\overline{\ell^c_{kL}} \, \ell_{mL} \ , 
\nonumber \\
O_{\ell q} & = &   \overline{\ell_{i}} \Gamma_\ell \ell_{j} \  
\overline{q} \Gamma_q  q  \  , 
\eea 
along with their corresponding chiral analogues ($L \leftrightarrow
R$).  Since operators of the type $O_{\sigma}$ typically arise at the
one-loop level, we explicitly display the loop factor $1/(4 \pi)^2$.
On the other hand, in a large class of models, operators of the type
$O_{\ell}$ or $O_{\ell q}$ may arise from both tree level exchange of
heavy particles as well as loop effects.  With the above choices, the
leading contributions to the various $c_i$ are nominally of the same
size, typically the product of two Yukawa-like couplings or gauge
couplings (times flavor mixing matrices).

\begin{figure}[!ht]
\centering
\epsfig{figure=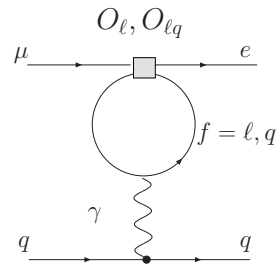,height=3.5cm}
\caption{
Loop contributions to $\mu \to e$ conversion through insertion of 
operators $O_{\ell}$ or $O_{\ell q}$, generating the large logarithm.
}
\label{fig:fig3}
\end{figure}

In terms of these operator definitions, the ratio ${\cal R}$ can be
written schematically as follows (neglecting flavor indices in the
effective couplings and the term with $L \leftrightarrow R$)
\bea 
{\cal R} &=& 
\displaystyle\frac{\Phi}{48 \pi^2} \,  
%\displaystyle\frac{ 
\Big| \eta_1  \, e^2  c_{\sigma L}  + e^2 \left( 
\eta_2   c_{\ell L} + \eta_3 c_{\ell q} \right)   
\log \displaystyle\frac{\Lambda^2}{m_\mu^2}
\nonumber \\
&+&   \eta_4 (4 \pi)^2  c_{\ell q} \ + \dots 
\Big|^2 / 
\left[ e^2 \left( |c_{\sigma L}|^2  + |c_{\sigma R}|^2 \right) \right] \, .
\label{eq:main1} 
\eea
Here, $\eta_{1,2,3,4}$ are numerical factors of $O(1)$, while the
overall factor $\Phi/48\pi^2$ arises from phase space and overlap
integrals involving electron and muon wave-functions in the nuclear
field. For light nuclei $\Phi = (Z F_p^2)/(g_V^2 + 3 g_A^2) \sim
O(1)$, where $g_{V,A}$ are the vector and axial nucleon form factors
at zero momentum transfer, while $F_p$ is the nuclear form factor at
$q^2 = -m_\mu^2$~\cite{Kitano:2002mt}.  The dots indicate sub-leading
terms, not relevant for our discussion, such as loop-induced
contributions to $c_{\ell}$ and $c_{\ell q}$ that are analytic in
external masses and momenta.  In contrast the 
logarithmically-enhanced loop contribution given by the second term in
the numerator of ${\cal R}$ plays an essential role. This term arises 
whenever the operators $O_{\ell L,R}$ and $O_{\ell q}$ appear at
tree-level in the effective theory and generate one-loop
renormalization of $O_{\ell q}$~\cite{Raidal:1997hq} (see
Fig.~\ref{fig:fig3}).

The ingredients in Eq.~(\ref{eq:main1}) lead to several observations:
(i) In absence of tree-level $c_{\ell L }$ and $c_{\ell q}$, one
obtains ${\cal R} \sim (\Phi \, \eta_1^2 \, \alpha)/(12 \pi) \sim
10^{-3}-10^{-2}$, due to gauge coupling and phase space
suppression. 
(ii) When present, the logarithmically enhanced contributions
compensate for the gauge coupling and phase space suppression, leading
to ${\cal R} \sim O(1)$. 
(iii) If present, the tree-level coupling $c_{\ell q}$ dominates the
$\mu \to e$ rate leading to ${\cal R} \gg 1$. 

In light of these observations, the logic underlying items 1 and 2 
can be phrased as follows.  In models with TeV scale $|\Delta L|=1,2$
interactions and generic flavor content of the 
couplings one finds $\tilde{c}_{i}/g^2 \sim c_{\ell L}, c_{\ell R},
c_{\ell q}$ ($g$ is a generic gauge coupling), and thus a
short-distance contribution to $0 \nu \beta \beta$ is accompanied by
${\cal R} \gg 10^{-2}$.  
We now illustrate this statement in two explicit cases: the
minimal supersymmetric standard model (MSSM) with R-parity violation
(RPV-SUSY) and the Left-Right Symmetric Model (LRSM).
\begin{figure}[!t]
\centering
\epsfig{figure=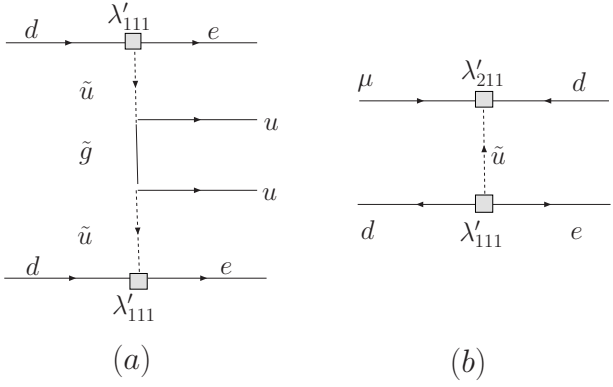,width=8.5cm}
\caption{
Gluino exchange contribution to $0 \nu \beta \beta$ $(a)$, 
and typical tree-level contribution to $O_{\ell q}$ $(b)$ in RPV SUSY. 
}
\label{fig:fig1}
\end{figure}

{\em RPV SUSY ---}  If one does not impose R-parity conservation [$R=
(-1)^{3 (B-L) + 2 s}$], the MSSM superpotential includes, in addition
to the standard Yukawa terms, lepton and baryon number violating
interactions, compactly written as (see e.g.~\cite{Dreiner:1997uz})  
\bea
W_{RPV} &=& \lambda_{ijk} L_i L_j E_k^c + 
\lambda_{ijk}' L_i Q_j D_k^c + 
\lambda_{ijk}''  U_i^c D_j^c D_k^c 
\nonumber \\
& & +  \mu_{i}' L_i  H_u   \ , 
\eea
where $L$ and $Q$ represent lepton and quark doublet superfields,
while $E^c$, $U^c$, $D^c$ are lepton and quark singlet superfields.
The simultaneous presence of $\lambda'$ and $\lambda ''$ couplings
would lead to an unacceptably large proton decay rate (for SUSY mass
scale $\Lambda_{SUSY} \sim$ TeV), so we focus on the case of
$\lambda'' = 0$ and set $\mu'=0$ without loss of
generality~\cite{Dreiner:1997uz}.  In this case, lepton number is
violated by the remaining terms in $W_{RPV}$, leading to short
distance contributions to $0 \nu \beta \beta$, with typical
coefficients [cf. Eq.~(\ref{eq:lag1})]
\beq 
\frac{\tilde{c_i}}{\Lambda^5} \sim 
\frac{\pi
\alpha_s}{m_{\tilde{g}}} \frac{\lambda_{111}'^2}{m_{\tilde{f}}^4} \, ; \, 
\frac{\pi
\alpha_2}{m_\chi} \frac{\lambda_{111}'^2}{m_{\tilde{f}}^4} \ ,  
\eeq
where $\alpha_s, \alpha_2$ represent the strong and weak gauge
coupling constants, repsectively.  The RPV interactions also lead to
lepton number conserving but lepton flavor violating operators [e.g.
 Fig.~\ref{fig:fig1}(b)], with coefficients
[cf. Eq.~(\ref{eq:lag2})]
\bea
\frac{c_{\ell}}{\Lambda^2} &\sim &
\frac{ \lambda_{i11} \lambda_{i21}^*}{m_{\tilde{\nu}_i}^2} , 
\frac{ \lambda_{i11}^* \lambda_{i12}}{m_{\tilde{\nu}_i}^2}  \ , 
\nonumber \\
%\quad 
\frac{c_{\ell q}}{\Lambda^2} & \sim & 
\frac{ \lambda_{11i}'^* \lambda_{21i}'}{m_{\tilde{d}_i}^2}, 
\frac{ \lambda_{1i1}'^* \lambda_{2i1}'}{m_{\tilde{u}_i}^2} \ , 
\nonumber \\
%\eeq 
%
%\beq
\frac{c_\sigma}{\Lambda^2} 
& \sim &  \frac{ \lambda \lambda^* }{m_{\tilde{\ell}}^2}, 
\frac{\lambda'  \lambda'^* }{m_{\tilde{q}}^2}  \ ,
%\qquad 
\eea
where the flavor combinations contributing to $c_{\sigma}$ can be
found in Ref.~\cite{deGouvea:2000cf}.  Hence, for generic flavor
structure of the couplings $\lambda$ and $\lambda'$ 
%(in particular, if
%the off-diagonal couplings of first and second generation are not
%extremely suppressed compared to the diagonal ones), 
%one sees that 
the underlying LNV dynamics generate both short distance
contributions to $0 \nu \beta \beta$ and LFV contributions that lead
to ${\cal R} \gg 10^{-2}$.  

Existing limits on rare processes strongly constrain combinations of
RPV couplings, assuming $\Lambda_{SUSY}$ is 
between a few hundred GeV and $\sim$ 1 TeV.  Non-observation of LFV at future
experiments MEG and MECO could be attributed either to a larger 
$\Lambda_{SUSY}$ ($>$ few TeV) or to suppression of couplings that
involve mixing among first and second generations.  In the former
scenario, the short distance contribution to $0\nu \beta 
\beta$ does not compete with the long distance one
[see Eq.~(\ref{eq:estimate})], so that $\Gamma_{0 \nu \beta \beta}
\sim m_{\beta \beta}^2$.  On the other hand, if the $\lambda$ and
$\lambda'$ matrices are nearly flavor diagonal, the exchange of
superpartners may still make non-negligible contributions to $0\nu
\beta \beta$.

{\em LRSM ---} The LRSM provides a natural scenario for introducing
non-sterile, right-handed neutrinos and Majorana
masses~\cite{Mohapatra:1979ia}.  The corresponding electroweak gauge
group $SU(2)_L \times SU(2)_R \times U(1)_{B-L}$, breaks down to
$SU(2)_L \times U(1)_Y$ at the scale $\Lambda \grts {\cal O}({\rm
TeV})$.  The symmetry breaking is implemented through an extended
Higgs sector, containing a bi-doublet $\Phi$ and two triplets
$\Delta_{L,R}$, whose leptonic couplings generate both Majorana
neutrino masses and LFV involving charged leptons:
%with quantum numbers
%$(1/2,1/2^*,0)$, $(1,0,-2)$, $(0,1,-2)$.  
%
%The physics of LFV is encoded in the leptonic Yukawa coupling, of 
%both Dirac and Majorana type, which by gauge invariance and parity read 
%
\bea
{\cal L}_Y^{\rm lept} &=& - \  
\overline{L_L}\, ^{i} \, 
\left( 
 y_D^{ij}  \,  \Phi \ + \   \tilde{y}_D^{ij} \,  \tilde{\Phi} 
\right) \, 
L_{R}^j  \\
&-&  
\overline{(L_{L})^c}\, ^i \    y_M^{ij} \, \tilde{\Delta}_L \  L_{L}^j
\ - \ 
 \overline{(L_{R})^c}\, ^i \  y_M^{ij}  \, \tilde{\Delta}_R  \ L_{R}^j \ . 
\nonumber 
\eea
Here $ \tilde{\Phi} = \sigma_2 \Phi^* \sigma_2$,
$\tilde{\Delta}_{L,R} = i \sigma_2 \Delta_{L,R}$, and leptons belong 
to two isospin doublets $L_{L,R}^i = (\nu_{L,R}^i,
\ell_{L,R}^i)$.  The gauge symmetry is broken through the VEVs
$\langle \Delta^0_R \rangle = v_R$, $\langle \Delta^0_L \rangle = 0$,
$ \langle \Phi \rangle = {\rm diag}(\kappa_1, \kappa_2) $. 
After diagonalization of the lepton mass matrices, LFV arises from
both non-diagonal gauge interactions and the Higgs Yukawa
couplings. In particular, the $\Delta_{L,R}$-lepton interactions are
not suppressed by lepton masses and have the structure ${\cal L} \sim
\Delta_{L,R}^{++} \, \overline{\ell_i^c} \, h_{ij} \, (1 \pm \gamma_5)
\ell_j + {\rm h.c.}$. The couplings $h_{ij}$ are in general
non-diagonal and related to the heavy neutrino mixing
matrix~\cite{Cirigliano:2004mv}.

\begin{figure}[!t]
\centering
\epsfig{figure=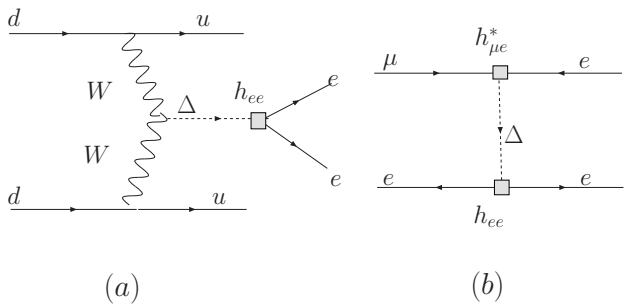,width=8.5cm}
\caption{
Typical doubly charged Higgs contribution to $0 \nu \beta \beta$ $(a)$ 
and to $O_{\ell}$ $(b)$ in the LRSM. 
}
\label{fig:fig2}
\end{figure}

Short distance contributions to $0\nu \beta \beta$ arise from the
exchange of both heavy $\nu$s and $\Delta_{L,R}$ 
(Fig.~\ref{fig:fig2}), with
\beq
\frac{\tilde{c}_i}{\Lambda^5} \sim  
\frac{g_2^4}{M_{W_R}^4} \frac{1}{M_{\nu_R}} 
\,  ; \, \frac{g_2^3}{M_{W_R}^3} \frac{h_{ee}}{M_\Delta^2} \ , 
\eeq
where $g_2$ is the weak gauge coupling. LFV operators are also generated 
through non-diagonal gauge and Higgs vertices, with~\cite{Cirigliano:2004mv}
\beq
\frac{c_{\ell}}{\Lambda^2} \sim \frac{h_{\mu i} h_{ie}^*}{m_{\Delta}^2} \qquad 
\frac{c_{\sigma}}{\Lambda^2} \sim 
\frac{(h^\dagger h)_{e \mu}}{M_{W_R}^2}  \quad   i=e, \mu, \tau \ . 
\eeq
Note that the Yukawa interactions needed for the Majorana neutrino
mass necessarily imply the presence of LNV and LFV couplings $h_{ij}$
and the corresponding LFV operator coefficients $c_{\ell}$, 
leading to ${\cal R} \sim O(1)$. 
Again, non-observation of LFV in the next generation of
experiments would typically push $\Lambda$ into the multi-TeV range,
thus implying a negligible short distance contribution to $0 \nu \beta
\beta$.  As with RPV-SUSY, this conclusion can be evaded by assuming a  
specific flavor structure, namely $y_M$ approximately diagonal 
or a nearly degenerate heavy neutrino spectrum.  

In both of the foregoing phenomenologically viable models
incorporating LNV and LFV at low scale ($\sim$ TeV), one finds
${\cal R} \gg 10^{-2}$~\cite{Raidal:1997hq,deGouvea:2000cf,Cirigliano:2004mv}.
We stress that the basic mechanism at work in 
these illustrative cases is likely to be generic: low scale LNV interactions
($\Delta L = \pm 1$ and/or $\Delta L= \pm 2$), which in general
contribute to $0 \nu \beta \beta$, also generate sizable contributions
to $\mu \to e$ conversion, thus enhancing this process over $\mu \to e
\gamma$.
Barring accidental cancellations or explicit fine-tuning of flavor
structures, this enhancement happens essentially through tree-level 
generation of four-fermion operators of the type $O_{\ell q}$ or
$O_{\ell}$.  

It is possible, of course, that any short-distance LNV sufficiently
strong to affect the interpretation of $\Gamma_{0\nu\beta\beta}$ is
also (approximately) flavor diagonal. In this case, one would expect
both branching ratios $B_{\mu\to e\gamma}$ and $B_{\mu\to e}$ to be
below the reach of up-coming LFV searches, and one would need
additional phenomenological handles to sort out the mechanism for a
non-zero signal in $0\nu\beta\beta$. Although seemingly un-natural,
such a situation is not precluded theoretically. Indeed, analogous
flavor fine-tuning in lepton number conserving models -- such as the
soft sector of the MSSM -- is not unheard of, and the phenomenological
absence of flavor changing neutral currents presents a challenge to
model builders to explain why a given theory is flavor diagonal when a
richer flavor structure is generically available. Were a similar
situation to arise in the presence of low-scale LNV, the theoretical
challenge would become all the more interesting. In this context, null
results for both LFV searches would provide important experimental
input.

In conclusion, we have argued that the ratio ${\cal R} = B_{\mu \to
e}/B_{\mu \rightarrow e \gamma}$ of muon LFV processes will provide
important insight about the mechanism of neutrinoless double beta
decay and the use of this process to determine the absolute scale of
neutrino mass.  Assuming observation of LFV processes in forthcoming
experiments, if ${\cal R} \sim 10^{-2}$ the mechanism of $0 \nu
\beta \beta$ is light Majorana neutrino exchange;  if ${\cal R}
\gg 10^{-2}$, there might be TeV scale LNV dynamics, and no definite
conclusion on the mechanism of $0 \nu \beta \beta$ can be drawn based
only on LFV processes.

\acknowledgments
This work was supported in part by the Institute for Nuclear Theory, 
U.S. DOE contracts \# DE-FG03-88ER40397 and 
\# DE-FG02-00ER41132,  and NSF Award PHY-0071856.  V.C. was
supported by a Sherman Fairchild Fellowship from Caltech. 
%\end{acknowledgments}

%  \vfill

\end{document}